\documentclass[aps, pre, showpacs, twocolumn, superscriptaddress,floatfix,]{revtex4}
\usepackage{amsmath,amssymb,graphicx,color,overpic}
\begin{document}

\title{Transitional steady states of exchange dynamics between finite quantum systems}
\author{Euijin Jeon}
\affiliation{Graduate School of Nanoscience and Technology, 
Korea Advanced Institute of Science and Technology, Deajeon 305-701, Korea}

\author{Juyeon Yi\footnote{corresponding author: jyi@pusan.ac.kr}}


\affiliation{Department of Physics, Pusan National University,
Busan 609-735, Korea}

\author{Yong Woon Kim\footnote{corresponding author: y.w.kim@kaist.ac.kr}}
\affiliation{Graduate School of Nanoscience and Technology, 
Korea Advanced Institute of Science and Technology, Deajeon 305-701, Korea}

\date{\today}

\hspace{2cm} \\

\begin{abstract}
We examine energy and particle exchange between finite-sized quantum systems and find a new form of nonequilibrium states.
The exchange rate undergoes stepwise evolution in time, and its magnitude and sign dramatically change according to system size differences. The origin lies in interference effects contributed by multiply scattered waves at system boundaries. Although such characteristics are
utterly different from those of true steady state for infinite systems, Onsager's reciprocal relation remains universally valid. 
\end{abstract}

\maketitle

\section{introduction}
One of the most fundamental phenomena in physics is energy and particle exchange between systems, which
occur in the form of heat and mass current in the presence of temperature and chemical potential gradient.
It is often our interest to understand steady state (SS) with constant exchange rate.  A great deal of research have been performed to clarify the properties of SS,
such as Landauer-B\"{u}ttiker (LB) formula for fermionic particle current~\cite{lb1,lb2,lb3,lb4}, 
Onsager's reciprocal relation in the linear response regime~\cite{onsager, casimir},
and thermoelectric effect~\cite{te1,te2,te3,te4,te5}. 
Also in modern formulation of stochastic thermodynamics, steady state fluctuation theorem explains the directionality of the flow as a consequence of the second law of thermodynamics~\cite{evans, gallavotti1, gallavotti2, esposito, talkner},
and proves symmetry relations between nonlinear response coefficients~\cite{saito1,saito2,jacquod} . 

In fact, decay from an initial transient state into SS usually occurs if a system is subject to a dissipation due to a coupling to an environment~\cite{Kampen, weiss}. Recent theoretical studies show that types of dissipation processes~\cite{xiong} and the presence of bound state~\cite{khosravi} are crucial for the formation of SS.  Numerical tools have been developed to examine how an open quantum system to reach SS~\cite{schiro}. 
It is worth noting that SS can also exist in a quantum system isolated from dissipative environment if the system itself is infinite to have continuum energy spectra. Elementary but illuminating example is Fermi's golden rule for the constant transition rate in a
system having continuous density of states~\cite{sakurai}.  

As for SS in isolated quantum systems, despite the very fact that quantum systems are not infinite in their size, 
we presume that if the energy levels of considered systems are spaced densely enough, SS would also be established in a very similar manner to infinite systems. In this regard, the assumption of infinite size or continuum energy levels seems only a matter of mathematical convenience.
However, it is obvious that the exchange rate between two finite systems cannot be constant perpetually. If so, we reach an unphysical situation, for example, that particles flow constantly from system A to system B even if system A is totally evacuated. Furthermore, finite quantum systems evolving according to time symmetric Schr\"{o}dinger equation cannot reach SS in the strict sense.
 Hence the behavior of SS predicted for infinite systems should cease to persist after a certain time scale $\tau_{c}$.  
The following questions arise. What determines $\tau_{c}$?  What are the subsequent states for a time longer than $\tau_{c}$?
Does any alternative form of SS emerge, which cannot be explained by existing theories for infinite systems? 
These are issues of fundamental importance in understanding exchange dynamics between isolated quantum systems.

We answer these questions for a minimal model composed of two systems of noninteracting fermions in one-dimensional chains.
The system details are introduced in Sec. II. In order to quantify exchange rate between the two systems, we consider particle and heat currents which are defined in Sec. III. 
We preform numerical calculation to obtain particle and heat currents between the two systems,
and results for the particle current are given in Sec. IV.  We then adopt a perturbative approach and get analytic results well explaining the numerical data, which is presented in Sec. V. The behavior of the heat currents and its implication to Onsager's reciprocal relation are
discussed in Sec. VI. Summary and discussion 
will follow as Sec. VII.  
 
\section{system}
We consider two fermonic systems (system $L$ and system $R$), each of which
is well described by a noninteracting tight binding Hamiltonian: 
\begin{equation}\label{ham}
{\cal H}_{\alpha}=-t\sum_{x_{\alpha}=1_{\alpha}}^{M_{\alpha}-2}(
c_{x_{\alpha}}^{\dagger}c_{x_{\alpha}+1}+c_{x_{\alpha}+1}^{\dagger}c_{x_{\alpha}})
\end{equation}
with $\alpha=L,R$. Here $c_{x_\alpha}$~($c_{x_\alpha}^{\dagger}$) symbolizes an operator which annihilates (creates) a fermionic particle at a site 
$x_\alpha$ in the system $\alpha$,
and satisfies anticommutation relations:
$
\{c_{x_{\alpha}},c_{x_{\alpha'}}\}=\{c^{\dagger}_{x_{\alpha}},c^{\dagger}_{x_{\alpha'}}\}=0, ~~
\{c^{\dagger}_{x_{\alpha}},c_{x_{\alpha'}}\}=\delta_{x_{\alpha},x_{\alpha'}}\delta_{\alpha,\alpha'}.
$
We focus on size effects, assuming that the two chains can be different only in their lengths.  
The model Hamiltonian ${\cal H}_{\alpha}$ describes various physical systems such as hard-core bosons in one-dimensional optical lattices~\cite{hcb}, quantum spin rotors~\cite{aa}, and naturally a system of electrons if spin degrees of freedom are irrelevant. 

Initially~(at time $\tau=0$), the system $\alpha$ is in grand canonical equilibrium state at the inverse temperature $\beta_{\alpha}$ and chemical potential $\mu_{\alpha}$. The corresponding initial density matrix reads as 
\begin{equation}\label{iniden}
\rho_{eq}=e^{-\beta_{L}({\cal H}_{L}-\mu_{L}{\cal N}_{L})}e^{-\beta_{R}({\cal H}_{R}-\mu_{R}{\cal N}_{R})}/({\cal Z}_{L} {\cal Z}_{R})~,
\end{equation}
where ${\cal Z}_{\alpha}$ is the grand canonical partition function of the system $\alpha$, and ${\cal N}_{\alpha}$ is an operator measuring the total number of particles in the system $\alpha$, ${\cal N}_{\alpha}\equiv \sum_{x_{\alpha}}{\cal N}_{x_{\alpha}}$ with the particle occupancy at $x_{\alpha}$, 
${\cal N}_{x_{\alpha}}= c_{x_{\alpha}}^{\dagger}c_{x_{\alpha}}$.
\begin{figure}[b]
	\centering
\includegraphics[width=0.85\linewidth]{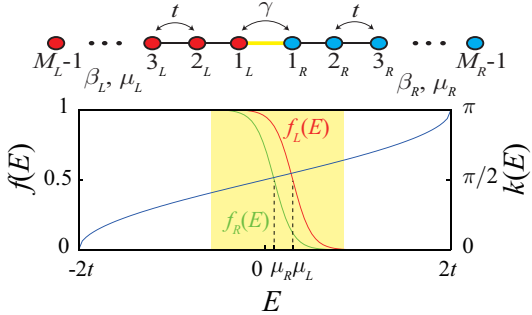}
	\caption{Schematic set-up: The upper figure shows the schematic diagram of the composite system, where $t$ and $\gamma$ represent the 
	hopping amplitudes defined in Eq.~(\ref{ham}) and Eq.~(\ref{coup}), respectively. There are $M_{\alpha}-1$ lattice sites in the chain $\alpha$, and 
	$\beta_{\alpha}, \mu_{\alpha}$  are parameters for the initial equilibrium states described by the density matrix (\ref{iniden}). 
	The lower figure exemplifies the Fermi-Dirac distributions,
	$f_{\alpha}(E)=[1+e^{\beta_{\alpha}(E-\mu_{\alpha})}]^{-1}$ with $\beta_{\alpha}$ and $\mu_{\alpha}$ satisfying the range (\ref{range}). Here the blue line 
	represents the energy dispersion, $E=-2t\cos k(E)$\cite{note}. Energy levels contributing to the currents populate the shaded region with 
	finite $f_{R}(E)-f_{L}(E)$, which is located near the band center $k(E)=\pi/2$. }
	\label{FIG1}
\end{figure}
A tunnel coupling between the two end sites (see Fig.~1) is switched on at $\tau=0^{+}$, which is described by a coupling Hamiltonian:
\begin{equation}\label{coup}
{\cal H}_{C}=-\gamma \left(c_{1_{L}}^{\dagger}c_{1_{R}}+c_{1_{R}}^{\dagger}c_{1_{L}}\right)~.
\end{equation}
The time evolution of the composite systems during $\tau >0$ is then governed by ${\cal H}\equiv {\cal H}_{L}+{\cal H}_{R}+{\cal H}_{C}$~\cite{chempot}.

There are several relevant energy scales in our consideration: $\beta_{\alpha}^{-1}$ and $\mu_{\alpha}$ for the initial equilibrium states, and the coupling strength $\gamma $.  In addition, for the system Hamiltonian Eq.~(\ref{ham}), we have the bandwidth $w=4t$ and the level spacing $\delta_{\alpha} \approx 2\pi t/M_{\alpha}$~\cite{note}. 
In this work, we consider weak coupling strength~($\gamma \ll t $), and highlight behaviors in a regime specified as
\begin{equation}\label{range}
\mu_{\alpha}, \beta_{\alpha}^{-1} \ll w, ~~\delta_{\alpha} \ll \beta^{-1}_{\alpha}.
\end{equation}
For the first condition, the systems are nearly half-filling and the contributions from the band edges are insignificant.  
For temperatures not much higher than room temperature, it is always $\beta_{\alpha} w \gg 1$ for an eligible range of band-widths, 
${\cal O}(10^{-1}) \mbox{eV} \lesssim w \lesssim {\cal O}(1)\mbox{eV}$.  The second condition requires that the systems should be large to have small level spacing compared to the thermal energy; hence, the level discreteness is irrelevant as for their initial equilibrium properties.  

\section{particle and heat currents}
In order to quantify exchange, we consider particle number change in the system $L$ :
\[
\left\langle \Delta\mathcal{N}_{L}(\tau)\right\rangle =\left\langle \mathcal{N}_{L}(\tau)\right\rangle -\left\langle \mathcal{N}_{L}\right\rangle.
\] 
The angular bracket of an observable ${\cal O}$ represents $\langle {\cal O} \rangle \equiv\mathrm{Tr}\rho_{eq}{\cal O}$ with $\rho_{eq}$ in Eq.~(\ref{iniden}). The operator of an observable ${\cal O}$ at time $\tau$ is represented by ${\cal O}(\tau)$ which is determined by the unitary time evolution:
$
{\cal O}(\tau)=U^{\dagger}(\tau) {\cal O}U(\tau) 
$
with $U(\tau)=e^{-i{\cal H}\tau/\hbar}$, and ${\cal O}(0)$ will be simply denoted as ${\cal O}$. 
Due to particle number conservation, the particle number change in the system $R$ is a redundant variable.
The particle current, $J_L(t)$, from the system $R$ to the system $L$ is then given by
\begin{equation}\label{JL}
J_N(\tau)=\langle d{\cal N}_{L}(\tau)/d\tau  \rangle~.
\end{equation}

We note here that a linear combination,
		\begin{equation}\label{basis}
		{\widetilde c}_{n_{\alpha}}=\sum_{x_{\alpha}=1}^{M_\alpha-1}a_{n_{\alpha},x_{\alpha}}c_{x_{\alpha}} 
		\end{equation}
		with the coefficients given by 
		\begin{equation}\label{decoupEigenstate}
		a_{n_{\alpha},x_{\alpha}}=\sqrt{\frac{2}{M_{\alpha}}}\sin \left(\frac{n_{\alpha}\pi x_{\alpha}}{M_\alpha}\right), 
		\end{equation}
		diagonalizes the Hamiltonian (1) into
		\begin{equation}\label{diagham}
		{\cal H}_{\alpha}=\sum_{n_{\alpha}}
		\varepsilon_{n_{\alpha}} {\widetilde c}^{\dagger}_{n_{\alpha}}{\widetilde c}_{n_{\alpha}} = \sum_{n_{\alpha}}
		\varepsilon_{n_{\alpha}} \widetilde{\cal N}_{n_{\alpha}}.
		\end{equation}
		Here $\widetilde{\cal N}_{n_{\alpha}}={\widetilde c}^{\dagger}_{n_{\alpha}}{\widetilde c}_{n_{\alpha}}$ is the number operator 
		measuring the number of fermion (0 or 1) occupying the $n_\alpha$-th energy eigenstate, and the energy eigenvalue is given as $\varepsilon_{n_{\alpha}}=-2t\cos (n_{\alpha}\pi/M_\alpha)$. In the diagonalizing basis, the coupling Hamiltonian $\mathcal{H}_{C}$ (3) is written as
		\[
		\mathcal{H}_{C}=\sum_{n_L,n_R}V_{n_L,n_R}\left(\widetilde{c}_{n_L}^{\dagger}\widetilde{c}_{n_R}+\widetilde{c}^{\dagger}_{n_R}\widetilde{c}_{n_L}\right)
		\]
		with $V_{n_L,n_R}=-\gamma a_{n_L,1_L}a_{n_R,1_R}$.  
		
		The number operator, $\mathcal{N}_L (\tau)=\sum_{n_L}\widetilde{\mathcal{N}}_{n_L}(\tau)$ for $\tau >0$ evolves according to the Heisenberg equation of motion:
		\begin{eqnarray}\nonumber
		i\hbar d_\tau  \mathcal{N}_{L}(\tau)&=&[\mathcal{N}_{L}(\tau),\mathcal{H}]\\ \nonumber
		&=&\sum_{n_{L},n_R}V_{n_L,n_{R}}\left(\widetilde{c}_{n_L}^{\dagger}(\tau)\widetilde{c}_{n_R}(\tau)-\widetilde{c}^{\dagger}_{n_R}(\tau)\widetilde{c}_{n_L}(\tau)\right),
		\end{eqnarray}
		and thus the particle current (\ref{JL}) can be expressed as 
		\begin{equation}\label{JNcor}
		J_{N}(\tau)=\frac{2}{\hbar}\sum_{n_{L},n_{R}}V_{n_{L},n_{R}}\mathrm{Re}[i\langle \widetilde{c}_{n_{R}}^{\dagger}(\tau)\widetilde{c}_{n_{L}}(\tau)\rangle],
		\end{equation}
		where $\mathrm{Re}[X]$ denotes the real part of $X$.  

Meanwhile, energy exchange occurs in the form of heat which is defined as \cite{andrieux,jarzynski, komatsu, panasyuk}:
$Q(\tau)=\left(\Delta\mathcal{E}_{L}(\tau)-\Delta\mathcal{E}_{R}(\tau)\right)/2~$.
Here, the energy change stored in
the system $\alpha$ is $\Delta \mathcal{E}_{\alpha}(\tau)=\left\langle \Delta {\cal H}_{\alpha}(\tau) \right\rangle -\mu_{\alpha}\left\langle \Delta \mathcal{N}_{L}(\tau)\right\rangle$ with $\Delta\mathcal{H}_{\alpha}(\tau)={\cal H}_{\alpha}(\tau)-{\cal H}_{\alpha}$. In the diagonalizing basis (\ref{basis}),
the energy change in the system $\alpha$ is given by 
\begin{equation}\nonumber
		 \Delta \mathcal{E}_\alpha(\tau)=\sum_{n_\alpha}(\varepsilon_{n_\alpha}-\mu_\alpha)[\langle \tilde{c}^\dagger_{n_\alpha}(\tau)\tilde{c}_{n_\alpha}(\tau)-\tilde{c}^\dagger_{n_\alpha}\tilde{c}_{n_\alpha}\rangle]~. 
		 \end{equation}
		 Using the Heisenberg equation of motion for 
		 $ \tilde{c}^\dagger_{n_\alpha}(\tau)\tilde{c}_{n_\alpha}(\tau)=\widetilde{\mathcal{N}}_{n_\alpha}(\tau)$, we obtain the time derivatives of the
		 energy change in the system $L$ as 
		 \begin{equation}\label{heatl}
		 d_\tau\Delta \mathcal{E}_L(\tau)=2\sum_{n_L,n_R}(\varepsilon_{n_L}-\mu_L)V_{n_L,n_R}\mathrm{Re}[i\langle c_{n_{R}}^{\dagger}(\tau)c_{n_{L}}(\tau)\rangle]
		 \end{equation}
		 which upon interchanging the system indices as $L\leftrightarrow R$ gives the time derivative of the energy change in the system $R$. 
		 This determines the heat current through a relation, $J_{Q}(\tau)\equiv d_\tau Q(\tau)=[d_\tau \Delta \mathcal{E}_{L}(\tau)-d_{\tau}\Delta \mathcal{E}_{R}(\tau)]/2$.

 \begin{figure}[tbp]
	\centering
\includegraphics[width=0.85\linewidth]{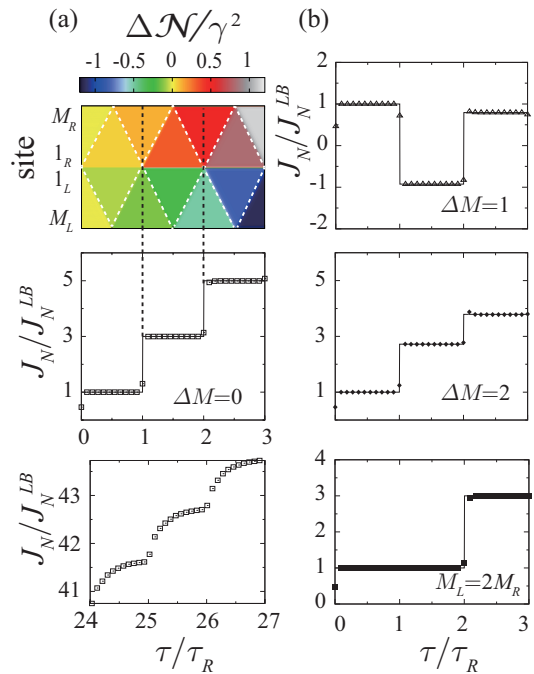}
	\caption{Time evolution of the particle currents:  In presenting the results, the current is normalized by $J^{LB}_{N}$, the current value 
	for infinite systems, which is obtained by the Landauer-B\"{u}ttiker formula. The time is scaled in units of $\tau_{R}=2M_{R}/v(0)$
	with $v(E)$ given by Eq.~(\ref{groupvel}), which is the minimum time required for a single round trip along the right system. Here, we set $\gamma=0.01 t$, $\beta_{L}=\beta_{R}=10/t$, $\mu_{L}=0.3 t$, and $\mu_{R}=0.1 t$.  (a) Results for a symmetric arrangement ($M_{L}=M_{R}=300$). The density plot of particle number change as a function of the position and the observation time (the upper panel), and the particle currents (the middle panel).  In the lower panel, we show the long time behavior where the step structure becomes vague because of the retardation of the round trip time of a slow particle (see the discussion in the paragraph below Eq.(\ref{groupvel})).  (b) The particle currents for $\Delta M\equiv  M_{L}-M_{R}=1$, $\Delta M = 2$ and $\Delta M=300$ with fixing $M_{R}=300$.} 
\end{figure}

\section{numerical results for particle currents}
We numerically calculate the currents and present the results for $J_{N}(\tau)$ in Fig.\,2 (Results for the heat currents will be discussed in Sec. VII). 
The upper panel of Fig.\,2\,(a) displays the density plot of the particle number variance, $\Delta {\cal N}_{x_{\alpha}}(\tau)$, for $M_{L}=M_{R}$.
The density variation propagates with the maximum velocity (the white dashed line) given by $v(0)$ in Eq.~(\ref{groupvel}) below, and it forms the triangular pattern. 
The middle panel of Fig.\,2\,(a) shows the particle current as a function of time, which evolves stepwise in time,
and the step heights are given by odd integer multiples of $J_{N}^{LB}$. Here, $J_{N}^{LB}$ is the steady state current for $M_{L}=M_{R}=\infty$, and it is
obtained from the Landauer-B\"{u}ttiker formula (see Appendix A). 
Also we scale observation time $\tau$ in units of $\tau_{R}$ with $\tau_{\alpha} \equiv 2M_{\alpha}/v(0)$, where the group velocity $v(E)$ is 
\begin{equation}\label{groupvel}
\hbar v(E)=\partial E(k)/\partial k=\sqrt{4t^{2}-E^{2}}~
\end{equation}
for the energy dispersion $E(k)=-2t\cos k$. Hence, the time scale $\tau_{R}$ corresponds to the shortest roundtrip time of a particle occupying the
band center ($E=0$) along the system $R$.  One finds that the currents abruptly jump at every time $\tau_{R} \equiv 2M_{R}/v(0)$.

The round trip time of a particle having energy $E$ is given by $\tau_{R} (E) \approx 2M_{R} [1+2(E/w)^{2}] /v(0)\equiv \tau_{R} [1+R(E)]$.  We find that the round trip of a particle with $E\neq 0$ is retarded to have a roundtrip time longer than $\tau_{R}$. 
This retardation determines the transient width between $m$th and $(m+1)$-th step, which is roughly given by $m\tau_{R}R(E)$. 
This effect can be observed in the current behavior for time $\tau=m\tau_{R}$ with large $m$, as illustrated in the lower panel of Fig.\,2\,(a). 
For a longer time, the step structure vanishes, and the current oscillates between positive and negative 
values~(see Appendix B).

We now look at other system size differences (Fig.\,2\, (b)). First note that, for $0< \tau \lesssim\tau_{R}$, the currents are given by $J_{N}^{LB}$, irrespectively of $\Delta M\equiv M_{L}-M_{R}$. This indicates that, up to the time $\tau_{R}$,  the systems do not sense their boundaries, and the exchange occurs in the same way as it does between infinite systems. However, as time elapses, the temporal behaviors of the currents can be very different from the symmetric case ($\Delta M=0$). 
The current amplitude sensitively depends on the size arrangement and observation time. Intriguingly, for the case $\Delta M=1$,  the current direction is  negative of $J_{N}^{LB}$, indicating backflow from low to high chemical potential. In the next section,
we derive an analytic formula well describing the numerical results and explain the size dependence of the current behaviors.  

\section{analytic result}
In order to evaluate the particle current (\ref{JNcor}) analytically, we need to calculate the equal time correlator 
$\langle\widetilde{c}_{n_{R}}^{\dagger}(\tau)\widetilde{c}_{n_{L}}(\tau)\rangle$. This can be done through a perturbative
approach for the weak coupling strength $\gamma \ll t$, as will be explained in the following. Note first that the time evolution of $\widetilde{c}_{n_\alpha}(\tau)$ is determined by the Heisenberg equation of motion,
		\begin{eqnarray}\label{EOM}
		i\hbar d_{\tau}\widetilde{c}_{n_{\alpha}}(\tau)&=&[\widetilde{c}_{n_{\alpha}}(\tau),\mathcal{H}]\\
		&=&\varepsilon_{n_{\alpha}}\widetilde{c}_{n_{\alpha}}(\tau)+\sum_{n_{\alpha^{\prime}\neq\alpha}}V_{n_{\alpha},n_{\alpha^{\prime}}}\widetilde{c}_{n_{\alpha^{\prime}}}(\tau). \nonumber
		\end{eqnarray}
		We then obtain $\widetilde{c}_{n_\alpha}(\tau)$ up to the linear order in $V$ for the weak coupling:
		\begin{eqnarray}\label{c}
		\widetilde{c}_{n_\alpha}(\tau)&\simeq& e^{-i\varepsilon_{n_{\alpha}}\tau/\hbar}\widetilde{c}_{n_{\alpha}}\\
		&&-\sum_{n_{\alpha^{\prime}\neq\alpha}}F_{n_{\alpha},n_{\alpha^{\prime}}}(\tau)e^{-i\varepsilon_{n_{\alpha^{\prime}}}\tau/\hbar}V_{n_{\alpha},n_{\alpha^{\prime}}}\widetilde{c}_{n_{\alpha^{\prime}}}~,\nonumber
		\end{eqnarray}
		where the time dependent coefficient $F_{n_{\alpha},n_{\alpha^{\prime}}}(\tau)$ is defined as
		\begin{eqnarray}\nonumber
		F_{n_{\alpha},n_{\alpha^{\prime}}}(\tau)&=&\frac{i}{\hbar}\int_{0}^{\tau}d\tau^{\prime}e^{i(\varepsilon_{n_{\alpha}}-\varepsilon_{n_{\alpha^{\prime}}})\tau^{\prime}/\hbar}\\ \label{f}
		&=&\frac{e^{i(\varepsilon_{n_{\alpha}}-\varepsilon_{n_{\alpha^{\prime}}})\tau/\hbar}-1}{\varepsilon_{n_{\alpha}}-\varepsilon_{n_{\alpha^{\prime}}}}~.
		\end{eqnarray}
		Inserting Eq.~(\ref{c}) into the equal time correlator in Eq.~(\ref{JNcor}), we find
		\begin{equation}\label{eqtimecor}
		\langle\widetilde{c}_{n_{R}}^{\dagger}(\tau)\widetilde{c}_{n_{L}}(\tau)\rangle\simeq-F_{n_{L},n_{R}}(\tau)V_{n_{L},n_{R}}\left[f_{R}-f_{L}\right]~,
		\end{equation}
		where we used a symmetry relation, $F^*_{n_R,n_L}(\tau)=-F_{n_L,n_R}(\tau)$, and the initial equilibrium averages, 
		$\langle\widetilde{c}_{n_\alpha}^{\dagger}\widetilde{c}_{n_{\alpha^\prime}}\rangle =
		\delta_{\alpha,\alpha^\prime}\delta_{n_\alpha,n_{\alpha^\prime}}f_{\alpha} $ with the Fermi-Dirac distribution function, $f_{\alpha}=[1+e^{\beta_{\alpha}(\varepsilon_{n_{\alpha}}-\mu_{\alpha})}]^{-1}$~. 
		The particle current can then be readily obtained by inserting Eq.~(\ref{eqtimecor}) into Eq.~(\ref{JNcor}): 
		\begin{eqnarray}\nonumber
		J_{N}(\tau)&=&\frac{2}{\hbar}\sum_{n_{L},n_{R}}V^2_{n_{L},n_R}\mathrm{Im}F_{n_{L},n_{R}}(\tau)\left[f_{R}-f_{L}\right] \\ \label{JNsum}
		&=&\sum_{n_{R}}T_{LR}f_{R}-\sum_{n_{L}}T_{RL}f_{L}~.
		\end{eqnarray}

                Let us first consider the term, $\sum_{n_{R}}T_{LR}f_{R}\equiv J_{N}^{R\rightarrow L}$ in Eq.~(\ref{JNsum}), corresponding to the current from the system $R$ to the system $L$. 
                We find that in time domain $2\tau/\tau_L \ll (w/E)^{2}$ the current is determined by (see Appendix C for the calculation details)
                                                                  \begin{equation}\label{curr2text}
                  J_{N}^{R\rightarrow L}\approx\int \frac{dE}{h} T_{LR}(E,\tau)f_{R}(E),
                  \end{equation}
                  where the transmission amplitude is given as
                  \begin{equation}\label{trantext}
                  T_{LR}(E,\tau)=T(E)\{1+2D(\tau)\sum_{\ell =1 }^{m}\cos[2k(E)C\ell]\}~.
                  \end{equation}
                  The time dependence of $T_{LR}$ lies in the factor, 
                  \begin{equation}\label{interval}
                  D(\tau)=\Theta(\tau-mq\tau_{R})\Theta((m+1)q\tau_{R}-\tau)
                  \end{equation}
                  with $\Theta(x)$ being the Heaviside step function and $q$ being a positive integer related to the system sizes as $p M_{L} = qM_{R} +C$ for its co-prime pair $p$ to make $C\sim {\cal O}(1)$. 
                  Here $T(E)$ is the transmission amplitude between semi finite chains, given in Eq.~(\ref{LBform}).
                  Exchanging the system indices, $L$ and $R$ in Eq.~(\ref{curr2text}) (and correspondingly, $q$ and $\tau_{R}$ in Eq.~(\ref{interval}) should be replaced with $p$ and $\tau_{L}$, respectively), we obtain 
                  the current from the left to the right system. Noting that $T_{RL}(E,\tau)$ is invariance under $L\leftrightarrow R$ and $p\leftrightarrow q$
                  due to $p\tau_{L}\approx q\tau_{R}$, we arrive at the following expression of the total current:
                                                      \begin{equation}\label{curr4text}
                  J_{N}=\int \frac{dE}{h} T_{LR}(E,\tau)(f_{R}-f_{L}),
                  \end{equation}
                  which is similar to the Landauer-B\"{u}ttiker formula, saving the fact that the transmutation amplitude is time dependent. 

The analytic results for the particle currents are represented by the lines in Fig.\,2, and they show good agreement with the numerical data. 
Let us briefly explain how the formula Eq.~(\ref{curr4text}) together with Eqs.~(\ref{trantext}) and (\ref{interval}) explain the numerical results. 
For example, if $M_{L}=M_{R}$, the time interval Eq.~(\ref{interval}) indicates that transmission jumps occur at every $\tau_{R}$,
as indeed displayed in the middle panel of Fig.\,2 (a). For this case, we have $C=0$ in Eq.~(\ref{trantext}), and the transmission amplitude for time interval
$m=1$ becomes $T_{LR}=3T(E)$, explaining the current value quantized at three times of $J_{N}^{LB}$.  On the other hand, if $\Delta M =1$,
we have $C=1$ and the cosine factor near the band center $k(E) \approx \pi/2$ becomes $\cos (\pi \ell)$, which for $\ell=1$ yields $T_{LR}=-T(E)$.
This negative transmission yields the negative current in the time interval given by $m=1$, as shown in the upper panel of Fig.2 (b). 
Behaviors for other cases can be deduced along the same line of reasoning.  

Note that $C$ in Eq.(\ref{trantext}) corresponds to the path difference between the round trip distance along the left and along the right system.
This suggests that the deviation from the Ladauer-B\"{u}ttiker formula originates from interference effects contributed
by waves reflected at the system boundaries and returning back to the coupling region. 
Furthermore, the commensurability of the round trip times $p\tau_{L} \approx q \tau_{R}$ and the time interval in Eq.~(\ref{interval}) indicates that the interference effect manifest itself only when the round trip along one system is concurrent with the other.

\section{heat currents and Onsager reciprocal relation}
We now look at behaviors of the heat currents, considering first the analytic expression (\ref{heatl}) of the energy change rate in the system $L$.
		 Substituting the equal-time correlator Eq.~(\ref{eqtimecor}) into Eq.~(\ref{heatl}), we get
		 \begin{equation}\label{eq1}
		 d_\tau\Delta \mathcal{E}_L(\tau)=\sum_{n_{L},n_{R}}2V^2_{n_L,n_R}(\varepsilon_{n_L}-\mu_L)\mathrm{Im}F_{n_L,n_{R}}[f_{R}-f_{L}]~.
		 \end{equation}
		 Further using a relation,
		  \begin{equation}\nonumber
		 \varepsilon_{n_L}\mathrm{Im}F_{n_L,n_{R}}=\varepsilon_{n_{R}}\mathrm{Im}F_{n_L,n_{R}}+\sin[(\varepsilon_{n_L}-\varepsilon_{n_{R}})\tau]~
		 \end{equation}
		 given from Eq.~(\ref{f}), 
                  we can write Eq.~(\ref{eq1}) as
                  \begin{eqnarray}\nonumber
                  d_\tau\Delta \mathcal{E}_L(\tau) &=&  \sum_{n_{R}}f_{R}(\varepsilon_{n_{R}}-\mu_{L}) \sum_{n_{L}} 2V^2_{n_L,n_R}\mathrm{Im}F_{n_L,n_{R}} 
                  \\ \nonumber
                  &-&\sum_{n_{L}}f_{L} (\varepsilon_{n_{L}}-\mu_{L}) \sum_{n_{R}} 2V^2_{n_L,n_R}\mathrm{Im}F_{n_L,n_{R}} \\ \nonumber
                  &+& \sum_{n_{R}}f_{R} \sum_{n_{L}}2V^2_{n_L,n_R} \sin [(\varepsilon_{n_L}-\varepsilon_{n_{R}})\tau]~.
                   \end{eqnarray} 
                   The last term is given by $S_{i}$ and $C_{i}$ with $i > 0$ in Eq.~(\ref{sandc}), which are delta functions or derivatives of delta functions as shown in Eqs.~(\ref{s1andc1}) and (\ref{othersc}).  Neglecting the last term acting only instantaneously and using the definition of $T_{LR}$ and $T_{RL}$ in Eq.~(\ref{JNsum}), we obtain
                   \begin{equation}\nonumber
                   d_\tau\Delta \mathcal{E}_L(\tau) = \sum_{n_{R}}f_{R}(\varepsilon_{n_{R}}-\mu_{L}) T_{LR} 
                   -\sum_{n_{L}}f_{L} (\varepsilon_{n_{L}}-\mu_{L})T_{RL}~.
                   \end{equation}
                  Converting the summation into integration, we can express the energy change rate of the left system as
		 \begin{equation}\nonumber
		 d_\tau\Delta \mathcal{E}_L(\tau)=\int \frac{dE}{h}(E-\mu_L)T(E,\tau)[f_{R}-f_{L}]~
		 \end{equation}
		 and finally reach the analytic formula for the heat current $2J_{Q}(\tau)=  d_\tau \Delta \mathcal{E}_L(\tau) -  d_\tau \Delta \mathcal{E}_R(\tau)$:
		 \begin{equation}\label{heatcurr}
		 J_Q(\tau)=\int \frac{dE}{h}(E-\bar{\mu})T(E,\tau)[f_R-f_L]~,
		 \end{equation} 
		 where $\bar{\mu}\equiv(\mu_L+\mu_R)/2$, and $T(E,\tau)=T_{RL}=T_{LR}$ is given in Eq.~(\ref{trantext}).

\begin{figure}
\centering
\includegraphics[width=.85\linewidth]{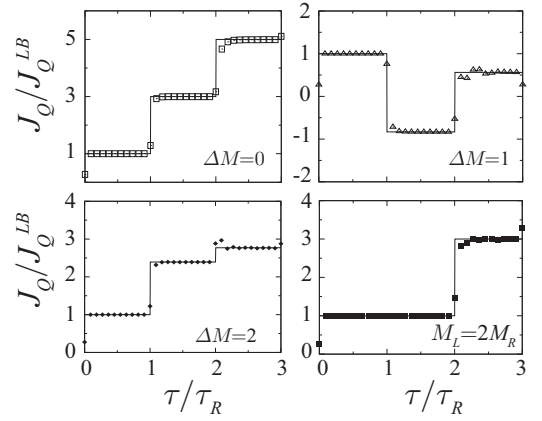}
\caption{Temporal behaviors of the heat currents for various $\Delta M$ with a fixed $M_{R}=300$. Here the heat currents
are normalized in units of $J^{LB}_{Q}$ given by Eq.~(\ref{infheat}),  and the observation time $\tau$ is scaled by $\tau_R$ as in Fig.~2. The parameters are set to be $\gamma=0.01t$, $\mu_L=\mu_R=\mu=0.2t$, $\beta_L=11/t$ and $\beta_R=9/t$.}
\label{Fig4}
\end{figure}
		
		In Fig.~\ref{Fig4}  we present the heat currents which are normalized by the steady state heat current,
		\begin{equation}\label{infheat}
		J_Q^{LB}= \int \frac{dE}{h} (E-\bar{\mu})T(E)[f_R-f_L]~.
		\end{equation}
		Analytic results (the lines) from Eq. (\ref{heatcurr}) are in good agreement with the numerical results (the points). 
		Also we can see that the heat currents evolve stepwise in time, similarly to the particle current behaviors shown in Fig.~2, 
		which could be understood from the time dependent transmission, $T(E,\tau)$. 		

The formula given in Eq.(\ref{curr4text}) for the particle current and in Eq.~(\ref{heatcurr}) for the heat current has an fundamental implication to the Onsager reciprocal relation~\cite{onsager,casimir}. 
Generally, particle currents can be expressed as $J_{N}(\tau)=\int dE (T_{LR}f_{R}-T_{RL}f_{L})$ with $T_{LR}$ ($T_{RL}$) denoting the transmission
amplitude from the right (left) to the left (right),  and $T_{LR}$ is not always equal to $T_{RL}$. 
The transmission in our consideration is shown to be symmetric under exchanging the system index $L$ and $R$, and there exists a symmetry,
\begin{equation}
\label{sym}
T_{RL}(E,\tau)=T_{LR}(E,\tau)=T(E,\tau),
\end{equation}
which leads to Eqs. (\ref{curr4text}) and (\ref{heatcurr}) in a form similar to the Landauer-B\"{u}ttiker formula. 

In the linear response regime, the particle current and the heat current can be approximated as $J_{N}\simeq L_{NN}\beta\Delta\mu+L_{NQ}\Delta\beta$ and 
$J_{Q}\simeq L_{QN}\beta\Delta\mu+L_{QQ}\Delta\beta$ for small affinity differences, $\Delta \beta = \beta_{L}-\beta_{R} \ll {\bar \beta}$ and $ \Delta \mu = \mu_{L}-\mu_{R} \ll {\bar \mu}$ with average temperature and chemical potential, $2{\bar \beta}=\beta_{L}+\beta_{R}$ and $2{\bar \mu}=\mu_{L}+\mu_{R}$ .
Expanding $f_{R}-f_{L}$ in Eqs.~(\ref{curr4text}) and (\ref{heatcurr}) up to the linear order in $\Delta \beta$ and $\Delta \mu$, 
one can readily check that those forms of the current formula, Eqs. (\ref{curr4text}) and (\ref{heatcurr}), although $T(E,\tau)$ is time dependent, validate the Onsager relation, $L_{NQ}=-L_{QN}$.
This is also confirmed by our numerical results shown in Fig.\,3. Therefore, Eq.~(\ref{sym}), a detailed balance condition can be viewed as the fundamental symmetry underlying the Onsager reciprocal relation.
\begin{figure}[t]
	\centering
        \includegraphics[width=0.85\linewidth]{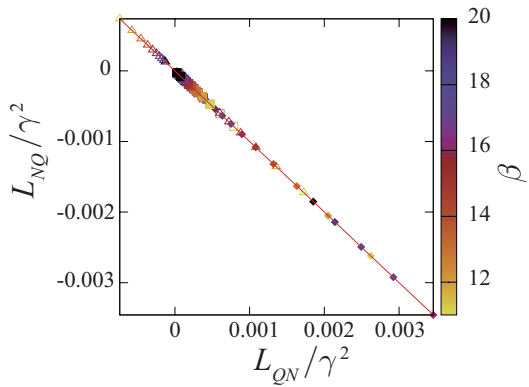}
	\caption{Onsager's reciprocal relation:
We evaluate $L_{NQ}$ and $L_{QN}$ for the cases presented in Fig.~2 at times $\tau= 0.5 \tau_{R}, 1.5 \tau_{R}$ and $2.5 \tau_{R}$, and all data points are collapsed onto the single line $L_{NQ}=-L_{QN}$. }
\end{figure}

\section{summary and discussion}
In this work, we suggest the existence of a new form of nonequilibrium state, characterizing exchange properties of finite-sized quantum systems.
A fundamental trait of the states is the stepwise evolution of currents with extreme sensitivity to system size difference, while still preserving the Onsager symmetry. Although we consider a one-dimensional fermonic system, underlying mechanisms is not restricted to the specific system, and similar size effects
must be present also for higher dimensional systems if their exchange dynamics are mainly governed by coherent (ballistic) transport. 

Experimental observation therefore depends on the availability of samples where particles maintain phase coherence. In this aspect, carbon nanotubes are a promising candidate material: Phase coherence of electrons is maintained over micrometers, and their conduction properties at low temperatures are well described by the theory of ballistic transport~\cite{cnt}.  
Also important is the time resolution of current measurement. For $w=0.1\sim 1$ eV and $M_{\alpha}=1\mu {\mbox m}/ 1\AA$, the round trip time is roughly estimated as 
$\tau_{\alpha}\approx 4(\hbar/w)M_{\alpha}\approx {\cal O}(10^{-1})\sim {\cal O}(10^{-2}) $ nsec. This gives a rough criterion for the required time resolution.
A superconducting quantum interference device (SQUID) can be most efficient for the current measurement, which detects magnetic fields 
generated by charge current flows with high sensitivity and picosecond time resolution~\cite{squid}. We do not answer how effects of particle interactions, interstitial defect and impurity modify the behaviors revealed here. In particular, at high temperatures electron-phonon scattering
must be a crucial phase-randomizing source (For carbon nanotubes the scattering time is about picoseconds at room temperature). These issues remain as important questions together with experimental challenges, which must be explored to advance our understanding of exchange phenomena in isolated quantum systems. 

\renewcommand{\thesection}{\Alph{section}}
\renewcommand{\theequation}{\Alph{section}.\arabic{equation}}
\setcounter{section}{1}

\appendix
\section{Landauer-B\"{u}ttiker formula}
The particle currents between infinite systems~($M_{L}=M_{R}=\infty$) can be obtained by using the Landauer B\"{u}ttiker formula, 
	\begin{equation}\label{lbcurr}
	J_{N}^{LB}=\int \frac{dE}{h} T(E) [f_{R}(E)-f_{L}(E)]
	\end{equation}
	with $h=2\pi \hbar$. 
	We consider the coupling sites, $1_{L}$ and
	$1_{R}$, as a small device connecting the two semi-infinite chains.  According to the transport theory~\cite{lb2,lb3,lb4}, 
	the transmission amplitude
	$T(E)$ is given by 
	 \begin{equation}\label{tran}
	T(E)=\Gamma^{2} |G_{1,2}|^2
	\end{equation}
	with $G(E)$ being the retarded Green function of the coupling device,
	\begin{equation}\label{green}
	G(E)=[E-{\cal H}_{C}-\Sigma]^{-1}=\left(\begin{matrix}
	E-\Sigma &\gamma\\
	\gamma &E-\Sigma
	\end{matrix}\right)^{-1}~.
	\end{equation}
       Here $\Sigma$ is the self energy for the coupling to the semi-infinite chains,
       \begin{equation}\label{SelfE}
	\Sigma = \frac{E}{2}-i\sqrt{t^2-\frac{E^2}{4}},
	\end{equation} 
       and its imaginary part gives the coupling function $\Gamma$ in Eq.~(\ref{tran}) as $\Gamma = -2\mbox{Im} \Sigma$. Using Eqs. (\ref{green}) and (\ref{SelfE}), 
        one obtains 
        		\begin{equation}\label{LBform}
	T(E)=4(\gamma/t)^2\left[1-\left(\frac{E}{2t}\right)^2\right].
	\end{equation}
	Inserting this into Eq.~(\ref{lbcurr}), we can evaluate the particle current between two infinite chains, $J_{N}^{LB}$.
\section{long time behaviors}
\begin{figure}[htp]
		\centering
		\includegraphics[width=0.85\linewidth]{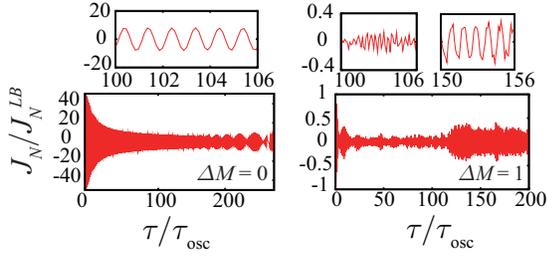}
		\caption{Particle current for $M_{L}=M_{R}$ and $M_{L}-M_{R}=1$ plotted for the long range of time, where we take the same parameters as used in producing Fig.~2 of the main text.  Here, time $\tau$ is scaled by $\tau_{osc}=hM/(4\gamma)$ for $M_{L}=M_{R}$ and by $\tau_{osc}=2\hbar M_{R}/t$ for $M_{L}-M_{R}=1$ (see the text). The upper panels are the enlarged views of the current oscillations. }
				\label{longtime1}
	\end{figure}

We present here particle current behaviors at longer times. Fig.~\ref{longtime1} shows $J_{N}$ as a function of time for $M_R=M_{L}=300$ in comparison with $M_{R}=300, M_{L}=301$. For the both cases, systems do not reach true steady state even in the long time limit. 
There are several points to be mentioned. Let us consider transitions between energy level in one system and its closest energy
level in the other system, which we let $\varepsilon_{n_{L}}$ and $\varepsilon_{n_{R}}$, respectively.  Due to the coupling described by ${\cal H}_{C}$ 
with its coupling strength $V_{n_{L},n_{R}} $ given by
\[
 V_{n_{L},n_{R}} = -\frac{2\gamma}{\sqrt{M_{L}M_{R}}} \sin (n_{L}\pi/M_{L})  \sin (n_{R}\pi/M_{R}) ,
 \]
 the two energy levels are hybridized into new levels having energies,
\[
E_{\pm}=\frac{1}{2}[\varepsilon_{n_{L}}+\varepsilon_{n_{R}}\pm 
\sqrt{(\varepsilon_{n_{L}}-\varepsilon_{n_{R}})^{2}+4|V_{n_{L},n_{R}}|^{2}}]~.
\]
Transition between the energy levels $E_{\pm}$ yields oscillation with period $\tau_{osc}=h/(E_{+}-E_{-})$. 
For the symmetric case ($M_{L}=M_{R}=M$), $\varepsilon_{n_{L}}=\varepsilon_{n_{R}}$, and the oscillation period is given by 
$\tau_{osc}=h/(2V_{n_{L},n_{R}})$ which for levels at the band center, that is, $n_{L}=n_{R}=M/2$, becomes $\tau_{osc}=hM/(4\gamma)$. As shown in the left upper panel,
the current oscillates with period $\tau_{osc}$, and also in the time domain not shown in the figure the oscillation period remains roughly $\tau_{osc}$.
This indicates that the rapid oscillation results from a resonant transition between two energy levels having same energy at band center. On the other hand, for the asymmetric case ($M_{R}=300, M_{L}=301$), we have
$\varepsilon_{n_{L}}- \varepsilon_{n_{R}} \gg 2 |V_{n_{L},n_{R}}| $ for the energy levels near the band center, and the oscillation period is determined by
the spacing between the unperturbed energy levels: $\tau_{osc}\approx h/(\varepsilon_{n_{L}}-\varepsilon_{n_{R}})\approx 2\hbar M_{R}/t$. 
Unlike the symmetric case, $\tau_{osc}$ only roughly fits the oscillation period during certain time intervals, for example, the time range of the upper right panel, and very noisy signals are present, as can be seen in the upper middle panel. 
Size effect appears not only in the rapid oscillation but also in the long time-scale behaviors. For the symmetric case, the amplitude decays inversely proportional to $\sqrt{\tau}$, and around $\tau=150 \tau_{osc}$ beating effect comes in. As time elapses, the beating frequency increases and the effect becomes more pronounced. For the asymmetric case ($M_L-M_R=1$), current behaviors are very distinctive from the symmetry case. 
Amplitude decay is accompanied by weak beating effect, and near $\tau=120\tau_{osc}$ large-amplitude periodic oscillation sets in.  Detailed analysis of these size effects in long time behaviors will be done in our future study.

\section{evaluation of $J_{N}^{R\rightarrow L}$}

		Let us first examine $T_{LR}$, the transmission amplitude from the right to the left system.  We write its explicit form, 
		\begin{eqnarray}\label{tmr}
		T_{LR}&=&\frac{2}{\hbar}\sum_{n_{L}}V_{n_{L},n_{R}}^{2}\mbox{Im}F_{n_{L},n_{R}} ,
		\end{eqnarray}
		where the wave numbers $k_{R}$ and $k_{L}$ are related to the energy level indices, $n_{L}$ and $n_{R}$, 
		as 
		\[
		k_{\alpha}=n_{\alpha}\pi/M_{\alpha}~,
		\]
		and $\mathrm{Im}F_{n_{L},n_{R}}(\tau)$ in Eq.~(\ref{f}) is 
				\begin{equation}\label{sindepde}
		\mathrm{Im}F_{n_{L},n_{R}}(\tau)=\frac{\sin[(\varepsilon_{n_L}-\varepsilon_{n_R})\tau/\hbar]}{\varepsilon_{n_L}-\varepsilon_{n_R}}~.
		\end{equation}
		From the factor $\mathrm{Im}F_{n_{L},n_{R}}$, we can see that transitions between adjacent energy levels, $\varepsilon_{n_{L}}\approx \varepsilon_{n_{R}}$,
		are dominant. Furthermore, since we are interested in physics coming from the band center where the energy levels are approximately linear
		 in $k_{\alpha}$, we find that $\Delta \equiv k_{R}-\pi/2$ and $x\equiv k_{L}-k_{R}$ are expansion parameters. The energy difference
		 in $\mathrm{Im}F_{n_{L},n_{R}}$ is expanded as
		 \begin{eqnarray}\nonumber 
		\varepsilon_{n_L}-\varepsilon_{n_R}&=& 2t\cos k_{R}-2t\cos k_{L} 
		\\ \nonumber
		& \approx & 2t x (1-\Delta^{2}/2 -\Delta x/2 -x^{2}/6) \\ \label{de3}
		&\equiv & 2tx +\Gamma(x,\Delta),
		\end{eqnarray}
	       and the coefficient in front of $\mathrm{Im}F_{n_{L},n_{R}}$ in Eq.~(\ref{tmr}) has an approximate form, 
	       \begin{eqnarray} \nonumber
	       \frac{2}{\hbar} V_{n_{L},n_{R}}^{2} &=&\frac{8\gamma^{2}}{M_{R}M_{L}\hbar }\sin^{2} (k_{R})\sin^{2} (k_{L}) 
	       \\ \nonumber
	       &\approx& \frac{8\gamma^{2}}{M_{R}M_{L}\hbar}(1-x^2 -2 x\Delta-2\Delta^{2})~.
	       \end{eqnarray}
		Using these expansions, we can express $T_{LR}$ as
		  \begin{equation}
		T_{LR}=\frac{4\gamma^{2}}{t \hbar M_{R}}\sum_{i=0}^{\infty}\left[A_{i}S_{i}(\tau)+B_{i}C_{i}(\tau)\right]\label{TnR2}~,
		\end{equation} 
		where the time dependent functions, $S_{i}(\tau)$ and $C_{i}(\tau)$, are defined by
		\begin{eqnarray}\label{sandc}
		S_{i}(\tau)&=&\frac{1}{M_L }\sum_{n_L}x^{i-1}\sin(2tx\tau/\hbar)\label{Ti}\\
		C_{i}(\tau)&=&\frac{1}{M_L }\sum_{n_L}x^{i-1}\cos(2tx\tau/\hbar)\label{Ui}.
		\end{eqnarray}
                Here we give a few coefficients relevant to our analysis:
                \begin{eqnarray}
                A_{0}&\approx &1-3\Delta^{2}/2, ~~ B_{0}=0 \\ \nonumber
                A_{1}&\approx & -3\Delta/2, ~~ B_{1}=-t\tau \Delta^{2} A_{0}/\hbar~.
                \end{eqnarray}
                
		
		Let us now evaluate Eq. (\ref{tmr}) with $T_{LR}$ in Eq.~(\ref{TnR2}), where the summation should be performed over $n_{L}$ for a given 
		$n_{R}$. We consider an energy level $n_{L}^{*}$ in the left system, which has the closest energy to $\varepsilon_{n_{R}}$: 
		\begin{equation}\label{defxi}
		n^*_L=(M_L/M_{R}) n_{R}+\xi,
		\end{equation}
		where $\xi$ is a number whose absolute value is less then $1/2$.
		Then  $x=k_{L}-k_{R}$ becomes
		\begin{equation}\nonumber
		x=\pi n_{L}/M_{L}-\pi n_{R}/M_{R}=(n_L-n_L^*+\xi)\pi/M_L.
		\end{equation}
		Since in our consideration the number of levels close to $n_{L}^{*}$ is sufficiently large for the convergence of the summation, we can extend
		the finite summation interval of $n_L$ in Eqs.(\ref{Ti}) and (\ref{Ui}) to infinite, $-\infty \leq n\equiv n_{L}-n^{*}_{L} \leq \infty$:
		\begin{eqnarray}
		S_{i}(\tau)&=&(M_L)^{-i}\sum_{n=-\infty}^{\infty}\frac{\sin[2\pi(n+\xi)\tilde{\tau}_L]}{[(n+\xi)\pi]^{1-i}}\label{Ti2}\\
		C_{i}(\tau)&=&(M_L)^{-i}\sum_{n=-\infty}^{\infty}\frac{\cos[2\pi(n+\xi)\tilde{\tau}_L]}{[(n+\xi)\pi]^{1-i}}\label{Ui2}~,
		\end{eqnarray}
		where $\tilde{\tau}_{L}=\tau/\tau_{L}$ with $\tau_{L}\equiv \hbar M_{L}/t$ being the minimum round trip time along the left system. Note here that
		the round trip time of a particle with wavenumber $k$  is given by
		\begin{eqnarray} \label{timeandvel}
		\tau_{L}(k)=2M_{L}/v(k),
		\end{eqnarray} 
		where the velocity of the particle is $v(k)=|2t\sin k|/\hbar$, and $\tau_{L}\equiv \tau_{L}(\pi/2)$ is the round trip time 
		of the fastest particle having $k=\pi/2$.

		We can evaluate $S_{1}$ and $C_{1}$, which are  the imaginary and the real part of $\sum_{n}e^{2i(n+\xi)\pi\tilde{\tau}_{L}}=\sum_{\ell}
		e^{2\pi i\ell\xi }\delta(\tilde{\tau}_{L}-\ell)$, respectively:
		\begin{eqnarray}\label{s1andc1}
		S_{1}(\tau)&=&\frac{1}{M_L}\sum_{\ell}\sin(2\pi \ell\xi)\delta(\tilde{\tau}_{L}-\ell)~,\\ \nonumber
		C_{1}(\tau)&=&\frac{1}{M_L}\sum_{\ell}\cos(2\pi \ell\xi)\delta(\tilde{\tau}_{L}-\ell)~.
		\end{eqnarray} 
		In determining $S_{i}$ and $C_{i}$ with $i \neq 1$, we use recursion relations,
		\begin{eqnarray}\label{othersc}
		S_{i+1}(\tau)&=&-\frac{1}{2M_L}\partial_{\tilde{\tau}_{L}}C_{i}~,\\ \nonumber
		C_{i+1}(\tau)&=&\frac{1}{2M_L}\partial_{\tilde{\tau}_{L}}S_{i}~,
		\end{eqnarray}
		 which can be derived from Eqs.(\ref{Ti2}) and (\ref{Ui2}).  We obtain $S_{0}$ by integrating $C_{1}$ over $\tilde{\tau}_L$ as		 \begin{eqnarray}\label{so}
		 S_{0}(\tau)&=&M_{L}\int^{\tilde{\tau}_{L}}_{-\tilde{\tau}_L}d\tilde{\tau}_L^\prime C_{1}(\tilde{\tau}_{L}^{\prime})\\
		 &=&1+2D(\tau)\sum_{\ell=1}^{m}\cos 2\pi \ell \xi~,\label{T0}\\
		 D(\tau)&\equiv& \Theta(m+1-\tilde{\tau}_{L})\Theta(\tilde{\tau}_{L}-m)~.
		 \end{eqnarray}
		Time dependence lies in the factor $D(\tau)$ with $\Theta(x)$ being the Heaviside step function, and we find that the value of $S_{0}(\tau)$ jumps at times integer multiples of $\tau_{L}$. 
                On the other hand, in (\ref{TnR2}) $C_{0}(\tau)$ makes null contribution because of the vanishing coefficients $B_{0}$.  Other terms with $i \geq 1$ in Eq.~(\ref{TnR2}) are delta functions as given in Eq.~(\ref{s1andc1}), or derivative delta function because $S_{i>1}$ and $C_{i>1}$ are given by the derivative $S_1$ and $C_1$ with respect to $\tilde {\tau}_{L}$. Therefore, the time dependence of $S_{0}$ essentially explains the temporal behavior of the currents shown in the main text.    
                 
                 Word of caution should be given here. The contribution from $C_{1}(\tau)$ in Eq.~(\ref{TnR2}) is non negligible for large $\tau$ because of its associated coefficient $B_{1}$ linearly increasing in $\tau$. The term $B_{1}C_1$ is written as
		 \begin{equation}\label{B1U1}
		 B_{1}C_1=\frac{(v(k_R)-v(\pi/2))\tau}{2M_L}A_{0}\partial_{\tilde{\tau}_L}S_0(\tau)~,
		 \end{equation}
		 where $v(k)$ is defined in Eq.~(\ref{timeandvel}). Including $B_{1}C_{1}$ to Eq.~(\ref{TnR2}), we obtain  
		 \begin{eqnarray}\nonumber
		 T_{LR}/(4\gamma^{2}/M_{R}t)&\approx &A_{0}S_{0}(\tilde{\tau}_{L})+(\tilde{\tau}_L(k_R)-\tilde{\tau}_{L})A_{0}
		 \partial_{\tilde{\tau}_L}S_{0}(\tilde{\tau}_{L}) \\ \nonumber
		 &\simeq &A_{0}S_{0}(\tilde{\tau}_L(k_R)),
		 \end{eqnarray}
		 where $\tilde{\tau}_L(k_R)$ is defined by $\tilde{\tau}_L(k_R)\equiv v(k_{R})\tau/(2M_L)$.
		 This indicates that the time duration function
		 $D(\tau)$ in Eq.~(\ref{so}) has wavenumber dependence as 
		 \[
		 D(\tau)\equiv \Theta(m+1-\tilde{\tau}_{L}(k))\Theta(\tilde{\tau}_{L}(k)-m), 
		 \]
		 and as a consequence the variation of $S_{0}$ occurs over a range of width which around time $\tau =m\tau_{L}$ is given by 		 
		     \[
		     m[ \tau_{L}(k_{R})- \tau_{L}] \approx m M_{L} \Delta^{2}\hbar/(2t) \approx 2\tau (E/w)^{2} .
		     \] 
		     Here the energy is approximated as  $E(k)\approx 2t \Delta $ near $k_{R}=\pi/2$, and $w=4t$ is the energy band width.
		     Since $E/w$ is very small for $E$ in the relevant energy range (4), the above variation width can be visible in the long time regime 
		     $2\tau (E/w)^{2} \gg \tau_L$. Therefore, in time domain $2\tau/\tau_L \ll (w/E)^{2}$ keeping only $S_{0}$ term, we find the transmission amplitude,
                                  \begin{equation}\nonumber
                 T_{LR}=\frac{4\gamma^{2}}{t\hbar M_{R}} (1-3\Delta^{2}/2)S_{0}(\tau)\equiv Q(n_{R})S_{0}(\tau)~,
                 \end{equation}
                 with $\Delta=k_{R}-\pi/2=n_{R}\pi/M_{R}-\pi/2$, which gives the expression of $J_{N}^{R\rightarrow L}$ as                
                  \begin{eqnarray}\label{currrtol}
                 J_{N}^{R\rightarrow L}=\sum_{n_{R}}f_{R}Q(n_{R})\left[1+2D(\tau)\sum_{\ell=1}^{m}\cos(2\pi\ell\xi)\right].
                                                \end{eqnarray}
                                              
                 Let us take a look at the cosine factor in the above equation:
                 \begin{eqnarray}\label{cos}
                 \cos(2\pi\ell \xi)&=&\cos[2\pi l (M_{L}/M_{R}) n_{R}].
                 \end{eqnarray}
                 Relating the system sizes as $pM_{L}-qM_{R}=C$, where $p$ and $q$ are positive integers and mutually prime, making $C$ to be a constant of
                 the order of unity (for example, for $M_{L}=402$ and $M_{R}=300$, $p=3$ and $q=4$, which give $C=6$),  we can write  Eq.~(\ref{cos}) as
                 \begin{equation}\nonumber
                 \cos(2\pi\ell\xi)=\cos\left[2 \pi n_{R}\left(q+ \frac{C}{M_{R}}\right) \frac{\ell}{p}\right]~,
                 \end{equation} 
                 which oscillates with $n_{R}$ and the oscillation period depends on the size factors. 
                 The phase component associated with integer $q$ may cause rapid oscillations as $n_{R}$ varies, if $\ell(q/p)$ is not an integer,
                 while the phase with $C/M_{R} \ll 1$ is a slowly varying component.  When performing summation 
                 over $n_{R}$ in Eq.~(\ref{currrtol}), non vanishing contribution is made by only terms with $\ell =p\ell^{\prime}$ with $\ell^{\prime}=1,2,\cdots$
                 because $q$ and $p$ are mutually prime.  Considering this fact, one arrives at 
                 \begin{equation}\label{curr2}
                J_{N}^{R\rightarrow L} =\sum_{n_{R}}f_{R}Q(n_{R})\left[1+2D(\tau)\sum_{\ell^{\prime} =1}^{m'}\cos(2k_{R}C\ell^{\prime})\right] 
                \end{equation}
                with the time dependent factor $
                                 D(\tau) =\Theta(p(m^{\prime}+1)\tau_{L}-\tau)\Theta(\tau-m^{\prime}p\tau_{L})$.
                                 
                                  We now change the summation
                                  over $n_{R}$ into integration with respect to energy $E=-2t\cos k_{R}$. Upon using
                                  \begin{equation}\nonumber
                                 \sum_{n_{R}} \approx \frac{M_{R}}{\pi} \int dk_{R} =\frac{M_{R}}{\pi} \int dE \rho(E)
                                 \end{equation}
                                 with the density of state for the one dimensional chains,  
                                 $\rho(E)=|dk/dE|\approx [1+E^{2}/(8t^{2})]/(2t)$, 
                   Eq.~(\ref{curr2}) becomes
                  \begin{eqnarray}\nonumber
                  \sum_{n_{R}}T_{LR}f_{R}&\approx&\int \frac{dE}{h} T_{LR}(E,\tau)f_{R}(E) \\ \nonumber
                  T_{LR}(E,\tau)&=&T(E)\{1+2D(\tau)\sum_{\ell =1 }^{m}\cos[2k(E)C\ell]\}.
                  \end{eqnarray}
                  Here $T(E)$ is the transmission amplitude between infinite chains, given in Eq.~(\ref{LBform}).


\begin{thebibliography}{10}   
	
\bibitem{lb1}
R. Landauer, IBM J. Res. Dev. {\bf 1}, 223 (1957): R. Landauer, J. Math. Phys. {\bf 37}, 5259 (1996); 
M. B\"{u}ttiker, Y. Imry, R. Landauer, and S. Pinhas, Phys. Rev. B {\bf 31}, 6207 (1985).

\bibitem{lb2}
S. Datta, {\it Electronic Transport in Mesoscopic Systems} (Cambridge University Press, Cambridge,1999).
\bibitem{lb3}
D. S. Fisher and P. A. Lee, Phys. Rev. B {\bf 23}, R6851 (1981).
\bibitem{lb4}
Y. Meir and N. S. Wingreen, Phys. Rev. Lett. {\bf 68}, 2512 (1992).
\bibitem{onsager}
L. Onsager, Phys. Rev. {\bf 37}, 405 (1931); Phys. Rev. {\bf 38}, 2265 (1931).
\bibitem{casimir}
H.~B. G. Casimir, Rev. Mod. Phys. {\bf 17}, 343 (1945).
\bibitem{te1}
R. S. Whitney, Phys. Rev. B {\bf 91}, 115425 (2015);  Phys. Rev. Lett. {\bf 112}, 130601 (2014)
\bibitem{te2}
A.N. Jordan, B. Sothmann, R. Sanchez, and M. B\"{u}ttiker, Phys. Rev. B, {\bf 87}, 075312 (2013).
\bibitem{te3}
D. Sanchez and R. Lopez, Phys. Rev. Lett. {\bf 110}, 026804 (2013).
\bibitem{te4}
J. Meair, and Ph. Jacquod, J. Phys. Condens. Matter 25 082201, (2013).
\bibitem{te5}
S. Hershfield, K.A. Muttalib, and B.J. Nartowt, Phys. Rev. B, {\bf 88}, 085426 (2013).
\bibitem{evans}
D. J. Evans, E. G. D. Cohen, and G. P. Morriss, Phys. Rev. Lett. {\bf 71} 2401 (1993); (E) 71 3616.
\bibitem{gallavotti1}
G. Gallavotti and E. G. D. Cohen, Phys. Rev. Lett. {\bf 74}, 2694 (1995).
\bibitem{gallavotti2}
 G. Gallavotti and E. G. D. Cohen, J. Stat. Phys. {\bf 80}, 931 (1995).
 \bibitem{esposito}
M. Esposito, U. Harbolar, and S. Mukamel, Rev. Mod. Phys. {\bf 81}, 1665 (2009).
\bibitem{talkner} 
M. Campisi, P. H¨anggi, and P. Talkner, Rev. Mod. Phys.
83, 1653 (2011).
\bibitem{saito1}
K. Saito and A. Dhar, Phys. Rev. Lett. {\bf 99} 180601 (2007).
\bibitem{saito2}
K. Saito and Y. Utsumi, Phys. Rev. B {\bf 78} 115429 (2008).
\bibitem{jacquod}
P. Jacquod, R.~S. Whitney, J.  Meair, and M. B\"{u}ttiker, Phys. Rev. B {\bf 86}, 155118 (2012).

\bibitem{Kampen}
N. V. Kampen, {\it Stochastic Processes in Physics and Chemistry} (North-Holland 1992).
\bibitem{weiss}
U. Weiss, {\it Quantum Dissipative Systems}, 2nd Ed. (World Scientific, Singapore 1999).
\bibitem{xiong}
H.-N. Xiong, P.-Y. Lo, W.-M Zhang, D. H. Feng, and F. Nori, Sci. Rep. {\bf 5}, 13353 (2015).
\bibitem{khosravi}
E. Khosravi, G. Stefanucci, S. Kurth, and E.~K.~U.. Gross, Phys. Chem. Chem. Phys. {\bf 11}, 4535 (2009).
\bibitem{schiro}
M. Schir'{o} and M. Fabrizio, Phys. Rev. B {\bf 79}, 153302 (2009) and references therein. 
\bibitem{sakurai}
J.~J. Sakurai, {\it Modern Quantum Mechanics} (Addison and Wesley 1994).

\bibitem{hcb}
M.~A. Cazalilla, R. Citro, T. Giamarchi, E. Orignac, and M. Rigol, Rev. Mod. Phys. {\bf 83}, 1405 (2011) .

\bibitem{aa}
A. Auerbach, {\it Interacting Electrons and Quantum Magnetism}, Springer-Verlag New York (1992).

\bibitem{chempot}
For the present consideration, the initial chemical potential difference is defined only for the initial equilibrium state, and during particle exchange system of interest is assumed to be decoupled from thermal and particle reservoirs which provide the initial equilibrium states.  

\bibitem{note}
For the system Hamiltonian (1), the energy band is given by $E_{\alpha}(k)=-2t\cos k_{\alpha}$ with the wavenumber $k$ and letting the lattice constant unity.
The band width is $4t$. 
Here the wavenumber is quantized as $k_{\alpha} = n_{\alpha}\pi/M_{\alpha}$ with $n = 1,2, \cdots, M_{\alpha}-1$. 

\bibitem{andrieux}
D. Andrieux, P. Gaspard, T. Monnai, S. Tasaki, New. J. Phys. {\bf 11}, 043014 (2009).
\bibitem{jarzynski}
C. Jarzynski and D. K.~W\'{o}jcik, Phys. Rev. Lett. {\bf 92} 230602 (2004).
\bibitem{komatsu}
T. S. Komatsu, N. Nakagawa, S. Sasa, and H. Tasaki, Phys. Rev. Lett. {\bf 100}, 230602 (2008).
\bibitem{panasyuk} 
G. Y. Panasyuk, G. A. Levin, and K. L. Yerkes, Phys. Rev. E {\bf 86}, 021116 (2012).

		
\bibitem{cnt}J. Kong, E. Yenilmez, T.~W. Tombler, W. Kim, H. Dai, R. B. Laughlin, L. Liu, C. S. Jayanthi, and S. Y. Wu, Phys. Rev. Lett. {\bf 87}, 106801 (2001).	

\bibitem{squid}D. D. Awschalom, J. Warnock, and S. von Moln\'{a}r, Phys. Rev. Lett. {\bf 58}, 821 (1987).
\end{thebibliography}
\end{document}